\documentclass{elsart}
\usepackage{natbib}
\usepackage{graphicx}
\input epsf.sty
\begin{document}
\runauthor{A.Janiuk, P. T. \. Zycki and B. Czerny}
\begin{frontmatter}
\title{Disk/corona model: The transition to ADAF}

\author[CAMK]{B. Czerny\thanksref{kbn}}
\author[CAMK]{A. R\' o\. za\' nska\thanksref{kbn}}
\author[CAMK]{A. Janiuk\thanksref{kbn}}
\author[CAMK]{P.T. \. Zycki\thanksref{kbn}}
\address[CAMK]{N. Copernicus Astronomical Center, Bartycka 18, 00-716 Warsaw, Poland}
\thanks[kbn]{Partially supported by the Polish State Committee for 
Scientific Research, grant 2P03D018.16}

\begin{abstract}
We propose a model of the accretion flow onto a black hole consisting of the
accretion disk with an accreting two-temperature corona. The model is based on 
assumptions about the radiative and conductive energy exchange between the two
phases and the pressure equilibrium. The complete model is determined by the 
mass, the accretion rate, and the
viscosity parameter. We present the radial dependencies of parameters of 
such a two-phase flow, with advection in the 
corona and the disk/corona mass exchange due to evaporation/condensation 
included, and we determine the transition radius from a two-phase 
disk/corona 
accretion to a single-phase 
optically thin flow (ADAF) in the innermost part of the disk as a function of 
accretion rate. We identify the
NLS1 galaxies with objects accreting at a rate close to the Eddington 
accretion rate. The strong variability of these objects may be related to
the limit cycle behaviour expected in this luminosity range, as the disk,
unstable due to the dominance by the radiation pressure, oscillates between the
two stable branches: the advection-dominated optically thick branch
and the evaporation branch.

\end{abstract}

\end{frontmatter}

\section{Introduction}

Broad band spectra of Seyfert galaxies and X-ray binaries in their hard 
states are well described by models which consist of a standard optically 
thick disk, disrupted and replaced by an optically thin hot flow in the
innermost part (e.g. Loska \& Czerny 1997, Esin, McClintock \& Narayan 1997).
The question remains: which physical mechanism is responsible for such a
transition?

Meyer \& Meyer-Hoffmeister (1994) suggested in the context of CVs that the
accretion flow is basically a two-phase disk/corona flow, with the heat 
generation in the corona leading to gradual disk evaporation due to the 
conduction flux,
and finally, to the disappearance of the disk in the innermost part of the
flow for low accretion rates.

Here we follow this basic idea but we adjust it to the parameter range 
appropriate for accreting black holes. 

\section{Model}

The accreting corona is cooled partially by advection, and its temperature is
of order of the virial temperature. 
Since the virial temperature in accreting black holes is orders of magnitude 
higher than in the case when a white dwarf is the central object because of its 
large radius, we modify the assumptions of Meyer \& Meyer-Hoffmeister 
considerably. We describe the corona flow as a two-temperature medium (see
Witt et al. 1997 and the references therein). We
include the effect of Compton cooling in addition to bremsstrahlung 
cooling. Finally, instead of using the conductive 
flux at the disk/corona boundary, taken from solar physics, we consider
the equilibrium at this boundary, as in R\' o\. za\' nska \& Czerny (2000a), 
and we calculate the evaporation rate, thus 
allowing for both evaporation or condensation (i.e. the settlement of the 
coronal material onto a disk). The details of the approach are given in
R\' o\. za\' nska \& Czerny (2000b).

\begin{figure}
\epsfxsize=8.8cm \epsfbox{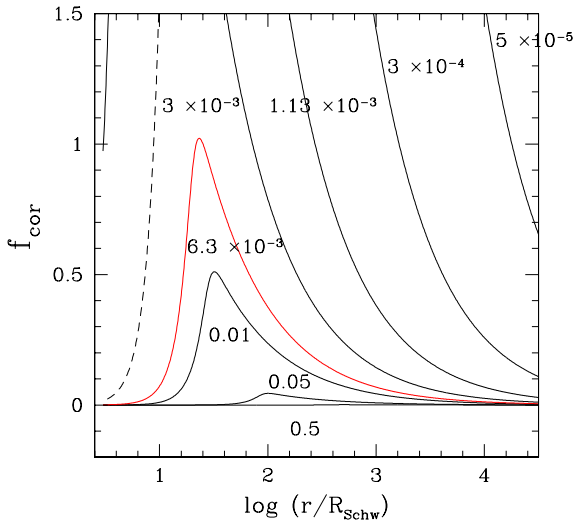}
\caption{The global solution for the radial dependence of the fraction of 
energy generated in the corona for $\alpha = 0.05$ and accretion rates
equal to $0.5$, $0.05$, $6.3 \times 10^{-3}$, $3\times 10^{-3}$, 
$1.13 \times 10^{-3}$, $3 \times 10^{-4}$, and $5 \times 10^{-5}$. The 
transition to ADAF is determined by $f_{\rm cor} = 1$. 
}
\label{fig:globseq}
\end{figure}

\section{Results}

An example of the radial dependence of the fraction of energy dissipated in 
the corona is shown in Fig.\ 1. For high accretion rate, the corona strength
reaches its maximum at some intermediate radius, condensation prevails in the
innermost part of the flow and the disk extends down to 
the marginally stable orbit. However, if the accretion rate is low, the 
evaporation is relatively efficient and the disk is replaced by the single 
phase hot flow at the radius where $f_{\rm cor} = 1$, i.e.\ the entire flow
is in the form of the hot phase and our solution is replaced by ADAF. 

The position of the transition radius is determined by the viscosity $\alpha$ 
and the accretion rate in Eddington units (with efficiency 1/12 included): 

\begin{equation}
\renewcommand{\arraystretch}{1.5}
\begin{array}{l@{\hspace{0.5cm}}l} 
r_{\rm tr}  =  4.51 \dot m^{-1.33} \alpha_{0.1}^7 R_{\rm Schw} & \dot m < 6.92 
\times 10^{-2} \alpha_{0.1}^{3.3}  \\
r_{\rm tr}  =  3  R_{\rm Schw}  &\dot m >  6.92\times 10^{-2}\alpha_{0.1}^{3.3} 
\end{array}
\renewcommand{\arraystretch}{1}
\label{por:tran}
\end{equation}
where $\alpha$ is expressed in 0.1 units.

For high accretion rates, the disk is surrounded by a relatively weak
corona and the radiation spectra are dominated by the disk component. However,
if the accretion is low, the disk evaporates and the spectrum is dominated by 
the ADAF component. Therefore, 
our model reproduces all characteristic luminosity states of an accreting black
hole without any additional {\it ad hoc} assumptions. 

The properties of the stationary models form a basis for the discussion of
the time evolution of the accretion flow, which is of interest because of the 
observed variability. 

The model stability is well represented by Fig.\ 2 where the positive slopes
mark stable solutions and negative slopes the unstable ones.
Since the disk at moderate accretion rates is surrounded by a weak corona, it 
remains unstable due to the radiation 
pressure dominance, as in the original model of 
Shakura \& Sunyaev. At still higher accretion rates, the optically thick 
disk is stabilized by advection while for lower accretion rates, we have a
strong departure from the classical models in the form of a new
evaporation branch,
which joins the optically thick solution to the ADAF solution.  

\begin{figure}
\epsfxsize=8.8cm \epsfbox{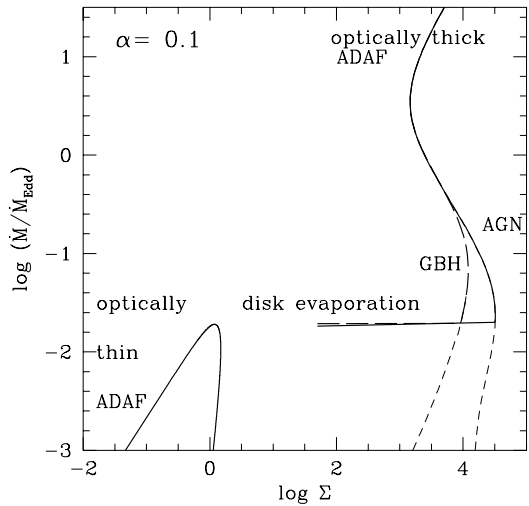}
\caption{The relation between the accretion rate and the surface density
of the disk/corona system at $10 R_{\rm Schw}$ for the viscosity parameter 
$\alpha = 0.1$ in the case of AGN ($M= 10^8 M_{\odot}$, continuous line) and  GBH ($M= 10 M_{\odot}$,
dashed line). 
Short-dashed lines show the corresponding standard Shakura-Sunyaev model.
The continuous line in the lower right part of the diagram shows an
exemplary solution for an optically thin flow  (Zdziarski 1998).}
\label{fig:mdotsig}
\end{figure}

Narrow Line Seyfert 1 galaxies, with their large accretion rates, correspond
to the intermediate, unstable branch at luminosities close to the 
Eddington value ($\dot m \sim 1$). According to our model, the accretion flow
may have a tendency to oscillate on a thermal and viscous timescale 
between the upper, disk dominated branch
and the horizontal evaporation branch with a strong, Comptonizing corona.
It may explain why those sources are generally strongly variable and it would
suggest that their counterparts among the Galactic sources are those in the
Very High State (and not High State, as frequently suggested).

\end{document}